# DeepRace: Finding Data Race Bugs via Deep Learning

Ali Tehrani
Computer Science Department
Iowa State University
Ames, IA, USA
tehrani@iastate.edu

Mohammed Khaleel
Computer Science Department
Iowa State University
Ames, IA, USA
mkhaleel@iastate.edu

Reza Akbari
IT & Computer Engineering
Shiraz University of Technology
Shiraz, Iran
akbari@sutech.ac.ir

Ali Jannesari
Computer Science Department
Iowa State University
Ames, IA, USA
jannesari@iastate.edu

## ABSTRACT

With the proliferation of multi-core hardware, parallel programs have become ubiquitous. These programs have their own type of bugs known as concurrency bugs and among them, data race bugs have been mostly in the focus of researchers over the past decades. In fact, detecting data races is a very challenging and important task. There have been several research paths in this area with many sophisticated tools designed and utilized that focus on detecting data race at the file level. In this paper, we propose DeepRace, a novel approach toward detecting data races in the source code. We build a deep neural network model to find data races instead of creating a data race detector manually. Our model uses a one-layer convolutional neural network (CNN) with different window size to find data races method. Then we adopt the class activation map function with global average pooling to extract the weights of the last convolutional layer and backpropagate it with the input source code to extract the line of codes with a data race. Thus, the DeepRace model can detect the data race bugs on a file and line of code level. In addition, we noticed that DeepRace successfully detects several buggy lines of code at different locations of the file. We tested the model with OpenMP and POSIX source code datasets which consist of more than 5000 and 8000 source code files respectively. We were able to successfully classify buggy source code files and achieve accuracies ranging from 81% and 86%. We also measured the performance of detecting and visualizing the data race at the line of code levels and our model achieved promising results. We only had a small number of false positives and false, ranging from 1 to 10. Furthermore, we used the intersection of union to measure the accuracy of the buggy lines of code, our model achieved promising results of 66 percent.

## 1 Introduction

With the ubiquity of multicore processors, the demand for multi-threaded programming that capable of fully using the power of modern systems is constantly on the rise. However, the prevalence of concurrent software has led to the emergence of bugs known as concurrency bugs. These bugs have caused some serious problems such as Northeast blackout [1] and NASDAQ's Facebook glitch [2]. Since these types of bugs are more common than ever in concurrent software [3], tackling them has been in the focus of researchers over the past years. One of the major sources of concurrency bugs is data race [4]. A data race bug occurs when two parallel accesses to one particular memory location are executed without an appropriate synchronization and at least one of the accesses is a write operation. It could often result in erroneous outputs [5].

Generally, data race detectors can be divided broadly into two categories. On one hand, there are static data race detectors [6]–[9] which typically examine source code or bytecode without executing the code. On the other hand, dynamic data race detectors [5], [10]–[13] observe and monitor the execution of the code either offline or online, we provide more details in the coming section.

With recent advancement in deep learning and the abundant source code available, researchers have been developing various deep learning models to tackle wide variety of software engineering and quality assurance tasks such as defect prediction [14], code clone detection [15] and code completion [16]. However, to the best of our knowledge, no significant work has been done on using deep learning to detect concurrency bugs, more specifically data race.

Figure 1 shows an example of a data race in a simple two-dimensional array computing program [17] which has one nested loop. Both variables $i$ and $j$ are declared outside of the parallel region but $i$, which is an index variable for the parallelized loop, is automatically private due to OpenMP implementation. The second loop, however, is not parallelized and its variable $j$ is shared among all threads. Therefore, in line 9, when read and write operations are executed, the value of $j$ could be anything assigned by any thread. This data race could be resolved by defining $j$ as a private variable to each. Consequently, each thread would have its own copy of variable $j$.

In this paper, we introduce DeepRace, a data race detector that addresses the problem of detecting data race bugs by leveraging the power of deep learning. For this purpose, we created a corpus of source code files from OpenMP and POSIX programs with specific synchronization patterns. Then, we built two classes of buggy and bug-free source code samples out of the corpus, i.e. a class for code samples that contain data races and another one for those files without data races. For simplicity will call the files with data race the buggy files and files without data race the bug-free files throughout the paper. We generated Abstract Syntax Trees (ASTs) for all the source code files. Then token vectors are created by extracting the class name of nodes from ASTs. Finally, we trained a deep learning model by feeding these vectors to the model.

Our first goal is to have a trained detector which can distinguish programs containing a data race from those without a data race. Unlike the previous works of deep learning and code analysis which classify source code files at method-level, our second goal is to further extend the approach to find the source of the bug at the line-level. To achieve this goal, we have adopted the class activates map method presented in [36] to extract the spatial location of the last convolutional layer neuron of the CNN classifier that reflects the location of the data race lines. We created three different datasets with three different data race bug patterns for OpenMP and POSIX programs. For this reason, we collected more than 15,800 source code files from GitHub and trained and test DeepRace with them. In addition, a test set of 60 sample programs with and without data race is created to compare DeepRace with state-of-the-art race detector tools to confirm the efficiency of our approach.

Our contributions in this work are summarized as follows:

- We implemented an efficient one-layer CNN classifier that applies various window size to detect the data race bugs related to three different synchronization patterns: *omp private clause*, *omp critical directives*, and *posix mutex locks*.

- While the recent works on code analysis with deep learning typically classify source-code at method-level, our model focus on detecting the data race at the line of code level. We achieved that by adapting the class activation map method to extract the weights of the various windows size convolutional layers and project it with the source code file to generate the line of codes that triggered the data race bugs. In addition, we noticed that DeepRace successfully detects several buggy lines of code at different locations at the file.

- With the lack of publicly available labeled source code dataset, we have collected dataset with three different data race bug patterns for OpenMP and POSIX programs. The dataset has 15,800 source code files, collected from GitHub. We equally divided the files to two categories: with and without data races, to address the problem of data imbalance that widely exists in the real-word datasets. We use the mutation method to generate source code files with data races. The dataset will be publicly available for researchers.[1]

The paper is structured as follows: In the next section (Section 2), we describe related works on data race bug detection and provide an overview of using deep learning models on other software engineering areas. Our approach is explained in Section 3. In Section 4, implementation details are outlined and discussed. The experimental results are presented in this section. Finally, in Section 5, we conclude the paper and outline the prospects of our future works.

## 2 Related Works

Many approaches of detecting the data races bugs exist in the literature. These approaches are commonly divided into two categories: dynamic and static detectors. For dynamic race detectors, *happens-before* and *lockset* algorithms are considered as a base for most of these tools [18]. A comprehensive description of these algorithms is provided in [12]. The static race detectors try to detect data races via analyzing source code without executing the code [19]. For our approach, since no code execution happens while analyzing the source code, we could consider DeepRace to be a kind of static detectors.

Recent advancements in machine learning and software analytics have led to address various software development challenges such as code completion and code clone detection. In the code completion, an intelligence feature in programming environments that helps to speed up the process of developing by suggesting fixes, Lie et al [16] have introduced an approach based on neural networks in particular LSTM [20]. LSTM is a type of neural network with state cells that act as long term and short term memory. Lie et al formulate the code completion problem as a sequential prediction task over the traversal of ASTs generated from JavaScript source codes. In [21] Reychev et al apply LSTM networks on a sequence of method calls to address the code completion for programs which are using APIs. Their work was later improved by introducing an approach based on decision tree learning [22]. Moving to the bug localization works, the procedure of locating buggy files considering a particular bug report, Lam et al [23] combine deep learning with an information retrieval technique called rVSM. Whereas in [24], since the characteristics of natural languages is different from that of programming languages, Huo et al applied two different convolution network architectures, considering the differences in the structure of programming language and natural languages. The intra-language features generated by convolution networks were fed to a fusion layer to indicate whether the source code file is related to a bug report or not.

Deep learning is also applied in code clone detection, the process of identifying duplicated code. Li et al [15] introduce CCLearner, an approach based on Deep Learning. It computes similarity vectors based on tokens for both clone pairs and non-clone pairs. These vectors are fed to the deep neural network in order to train a classifier to detect clone pairs and non-clone pairs. In the defect prediction context. Wang et al [25] applied Deep Belief Network [26] on token vectors which were generated by extracting tokens from source code's ASTs. Li et al [14] applied CNN on tokens vector combined by traditional features to improve the accuracy of file level defect prediction. Tingting Yu et al [37] for the first time proposed features for concurrency defect prediction. They

```c
int a[100][100];
int main()
{
  int i,j;
#pragma omp parallel for
  for (i=0;i<100;i++)
    for (j=0;j<100;j++)
      a[i][j]=a[i][j]+1;
  return 0;
}
```

```c
int a[100][100];
int main()
{
  int i,j;
#pragma omp parallel for private(j)
  for (i=0;i<100;i++)
    for (j=0;j<100;j++)
      a[i][j]=a[i][j]+1;
  return 0;
}
```

**Figure 1: An example of a data race in OpenMP program (left), resolving the data race via synchronization primitive (right)**

---

[1] Download link of the dataset will be shown in the final version of the paper

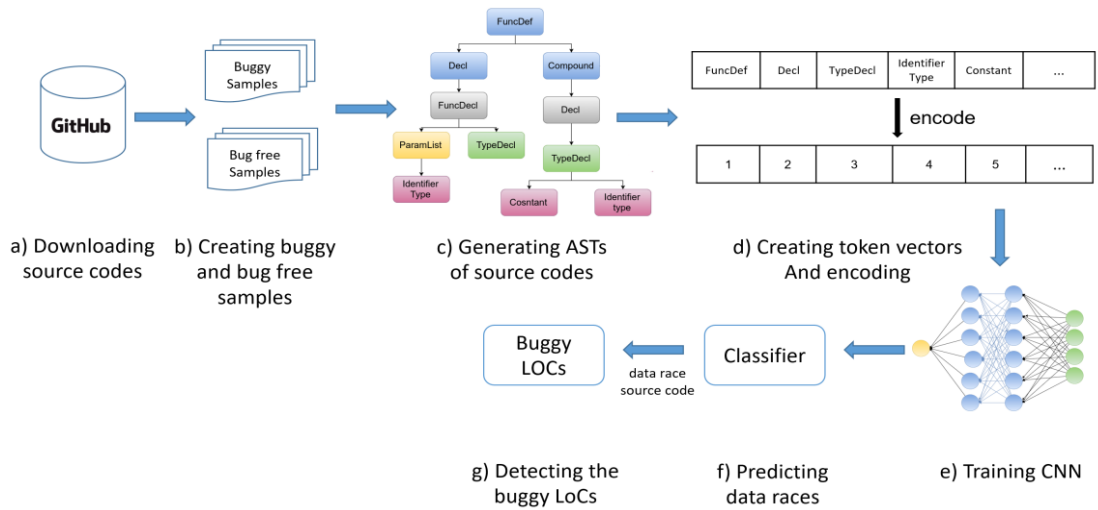

**Figure 2: The overview of DeepRace workflow**

proposed static and dynamic features in order to predict whether a method has defects or not. The prediction task was done using Machin Learning models namely: Bayesian Network, Decision Tree, Logistic Regression and Random Forest. The main drawback of their work is that the prediction is done only of method-level granularity meaning that it is not possible to predict which lines of code are involved in a concurrent defect.

While deep learning is applied to address various software development issues, to the best of our knowledge, no significant work has been done on detecting concurrency bugs specifically those which are related to data races.

## 3 Approach

In this section, we present the DeepRace, an approach built upon a CNN model which automatically learn structural and semantic information from programs and predict whether a program has an instance of data race pattern or not. The overall procedure of DeepRace is depicted in Figure 2.

We first collected a corpus containing OpenMP and POSIX source files with specific synchronization patterns from GitHub. The majority of these files are bug-free which leads to high data imbalance in terms of source code with and without data races. We address this problem by manually generating source code files with data races from the collected files for the training purposes using the mutation method presented in [38]. Next, the source code files are parsed, and an AST is generated for each file. Then, a vector containing nodes of AST is constructed which is fed to the following encode phase. In this phase, each token is embedded to a vector of a numerical values that is input to the CNN classifier. Semantic and structural features are automatically learned by the CNN. After training DeepRace, it would be able to probabilistically distinguish whether a source code contains data race or not given unseen source code. We also used the class activation map method [36] with the global average pooling to extract the weight of the neurons at the last convolutional layer and backpropagate it to the input layer to determine the line of codes that contain the data races. In the following sections, we explain each phase of our framework in details.

### 3.1 Collecting Data

The amount of training data used to train a deep learning model plays a crucial role in how well and accurate a model is trained. A number of datasets (consist of source codes and ASTs) exists in literature such as ETH JavaScript dataset [27] and Python source code dataset [28]. However, to the best of our knowledge, there is no publicly available dataset for the purpose of data race bug detection. Fortunately, many developers use a type of version control hosting service to keep track of their source codes and their corresponding versions. GitHub is known to be one of the most popular repositories hosting among developers. We implement a tool which utilizes GitHub's search feature and extracts source code files from open source projects available on GitHub. This tool, in particular, looks for source code files within particular synchronization patterns in it and downloads that file. These patterns are explained in the following subsections. In this research, the focus is on collecting files written in C which make use of parallel programming models OpenMP or POSIX threads. We leave collecting and training our model with more different types of concurrent bugs as future work. We believe the aforementioned concurrent bugs that cause the data races problem is very important to address in this paper. We collected an overall of 15,800 source code files written in C and created 3 datasets, namely OpenMP dataset 1, OpenMP dataset 2 and POSIX dataset.

**OpenMP Dataset 1**: In OpenMP, all variables are shared by default. Therefore, it is essential to declare a variable private (by using OpenMP private clause) to each thread where a concurrent write or read and write operation is performed on it. Missing or neglecting this clause could lead to a data race. Table 1 shows a total number of 5710 OpenMP programs with total 456,591 lines of code (LOC) are downloaded from GitHub. All of them make use of OpenMP private clause. 4568 of these files are used for training DeepRace while the rest are left for validation which means the split of roughly 80% / 20%.

**OpenMP Dataset 2**: For some synchronization scenarios a specific region of code has to be executed by only one thread at a time, this can be achieved via critical directive in OpenMP. Table 1 shows in total there are 1824 source code files in our OpenMP corpus, which consists 150,902 overall LOC, downloaded from projects available on GitHub. 1459 of those files are separated for training the model and validation is done on 365 files (80% / 20%).

**POSIX Dataset**: Mutexes is a way to protect shared variables. *Pthread_mutex_lock()* locks a mutex object and the thread calling this function becomes the owner of the mutex object until the same thread unlocks the object. If a thread tries to lock a mutex object which is already locked, that thread will wait until that object becomes available. A total of 8266 source code files containing *pthread_mutex_lock()* and *pthread_mutex_unlock()* are collected into our POSIX corpus. This includes 671,945 LOC totally. 6612 and 1654 files are used for training and validating the model respectively.

## 3.2 Creating Buggy and Bug-free Samples

Most of the collected source code files can be considered data race free since their source code style follow particular synchronization patterns (for OpenMP they either use private clause or critical directive and for POSIX programs they use POSIX lock primitive). To create the category for buggy files, we adopt the mutation generation method to generate data races source code by removing statements corresponding to synchronization primitives [38]. We apply the mutation generation method to the 50 percent of those source code files to generate buggy source code. In this way, we will have equal numbers of buggy samples (i.e. with data race) and bug-free samples (i.e. without data race). For OpenMP dataset 1, we intentionally injected a data race bug by removing the private clause from the source code using a regular expression (regex). Whereas for OpenMP dataset 2, buggy samples are generated by removing statements declaring critical sections and finally for POSIX dataset, those lines which are related to locking and unlocking mutex objects are removed. Since a data race is seeded only to half of the available files in datasets, therefore the data in both categories i.e. buggy and bug-free is balanced.

## 3.3 Parsing Source Code

The source code needs to be presented as vectors for the input of the neural network. This representation can be created on different degrees of granularity such as character level, token level, and nodes of AST. Mou Le et al [29] have shown that in order to keep both structural and syntactic information, using nodes of AST is a proper granularity. We follow their method as well. We deploy a C programming language parser Pycparser [30] and Clang [39] to generate the AST of the source code files for POSIX thread and OpenMP respectively. We have used the Clang parser since Pycparser is unable to parse OpenMP pragmas. It is worth mentioning that for POSIX programs ASTs are generated at file level but for OpenMP since a section of a method's body is usually parallelized, as a result, ASTs generated for OpenMP programs are at the method level. A token vector is generated by traversing the tree in depth first order and extracting the class name of AST's nodes (token types) such as `FuncDecl`, `TypeDecl`, `IdentifierType`, `Compound`, and `FuncCall` thus at the end of this phase for each source code we have a token vector. Figure 3 shows a part of AST of the Figure 1-left with its corresponding token vector (below AST). Since, generally the ASTs and their token vectors are so long, for the purpose of illustration, only a small part of the AST and its token vector is shown here. This AST is generated by utilizing Clang. The number of different token types for OpenMP dataset 1 and OpenMP dataset 2 are 209 and 151 respectively, but this number is 46 for POSIX dataset which probably indicates that using another parser might result in having more token types leading to having more information in the constructed dataset.

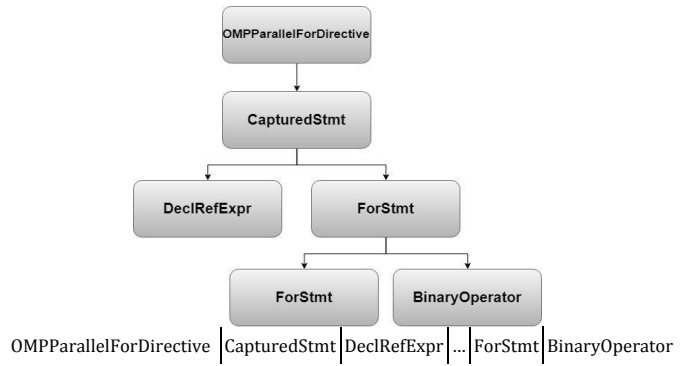

**Figure 3: A fraction of motivating example's AST and its corresponding token vector**

## 3.4 Encoding Token Vectors

Processing text and strings in almost all deep learning algorithms are not possible. Therefore, token vectors of a string type cannot be fed directly to CNN. To present tokens in integer, a mapping technique is implemented to map each token in a vector to a specific integer. The values of the integer are from 1 to the total number of token types. As a result, each token has a unique identifier. Also, a word embedding layer is employed. Word embedding represents a fixed dense vector which is the projection of a word in a continues vector space. A word's position in the vector space is learned from the context of that specific word in the input vector.

## 3.5 Convolutional Neural Network (CNN)

**Diverse vector size issue**: The source code files are diverse in terms of their length, resulting in vector embeddings of a wide length variety. To train a CNN model, the input training samples (token vectors) required to have the same length. One possible solution is to pad all vectors to the size of the longest vector. This might not be the most efficient way to tackle the problem, because some vectors might be too long, and this could waste the resources for handling dummy tokens. Another solution is to calculate the average length of all vectors, therefore vectors with length less than the average will be padded whereas those with longer length are shrunken by discarding exceeding tokens. We experimented with both models and observed that although padding all vectors to the longest vector consumes more resources and increases the training time, overall it helps to achieve better results, as it captures all the features information from the source code in contrast with the averaging solution some information might be discarded. Therefore, we chose the longest vector solution and padded all the vectors to the longest one. While Long-Short Term Memory (LSTM) would have the better characteristic of handling the various length input, the CNN ability to capture the spatial information makes it more suitable to capture the exact line of code which is causing the data races.

**Table 1: Statistics of three datasets**

| Datasets | Training files | Validation files | Total | Tokens | LOC |
|---|---|---|---|---|---|
| OpenMP dataset 1 | 4,568 | 1,142 | 5,710 | 209 | 456,591 |
| OpenMp dataset 2 | 1,459 | 365 | 1,824 | 151 | 150,902 |
| POSIX dataset | 6,612 | 1,654 | 8,266 | 46 | 671,945 |

**CNN architecture**: The proposed CNN model is implemented in Keras [31] backed by TensorFlow [32]. There are many complex deep learning architectures introduced in some theoretical approaches such as [33] and [34]. we employ the common existing architectures of CNN. A CNN model generally consists of several layers such as embedding, convolution, pooling, fully-connected and output. In our model, the first layer is the embedding which embeds the source code tokens into a dense vector of fixed-size. These dense vectors are then fed into the convolution layer, where different window size kernels are applied to find and detect important features of various code tokens.

The number of convolution and pooling layers are diverse and highly dependent on the training data and the field of study. A typical architecture in Figure 4 that is more prevalent in text classification field is that the output of embedding layer is usually fed to several parallel convolution layers each of which could be followed by a maximum pooling layer. The architecture and implementation details are provided in Table 2. After training, DeepRace will be able to probabilistically distinguish a file containing a specific data race from a bug-free file.

**Detecting the line-of-code data race**: What distinguished our model than the other presented in the literature is that DeepRace not only efficiently detect the buggy file but highlight the lines of code that causing the bugs. As a result, fixing the bugs of the source is time efficient and saving human effort trying to locate the error. To achieve that, we have adopt the class activation map method [36]. The class activation map function will simply identify the lines of code of the file that is being used by the CNN to classify a particular category. We first get the output of the softmax classification layer based on the classifier prediction. Then we do multiplication sum to the softmax layer weights with the values of the last convolutional layer of the CNN. Finally, we averaged the values of the several windows size layers and project it with the input source code to generate heat-map for each line of the source code.

More formally, given a source code file $x$, we need to visualize the lines of code that cause the data races problem based on the prediction $f(x) = 1$, where 1 indicate the classification of buggy file. Let $g_k(x_i)$ be the activation function output of neuron $k$ of the last convolutional layer. Then, we average the output values of neurons $k$ of all the $i^{th}$ filters at the last convolutional layer as follows:

$$H_k = \sum_i g_k(i)$$

We feed the average pooling value $H_k$ of the last convolutional layer to the softmax function to predict the category of the source code file as follows:

$$S_c = \frac{exp(\sum_k w_k^c H_k)}{\sum_c exp(\sum_k w_k^c H_k)}$$

Where $w_k$ is the weight of the softmax layer that directly connects the neuron $i$ of the convolutional network to class $c$. From the given equation, we notice that the score of class $c$ is highly related to the weight $w_k^c$ which represents the importance of $H_k$.

We generate the class activation map $G_c$ of class $c = 1$ at the spatial location i as follows:

$$G_c(i) = \sum_k w_k^c g_k(i)$$

Finally, we back-distribute the importance weights $G_c$ for each neuron $k$ of the convolutional network to the input layer that represents the source code file to detect the buggy line of the entire code.

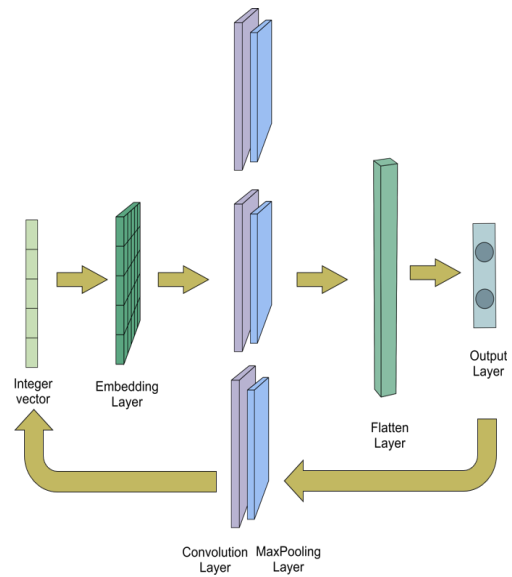

**Figure 4: DeepRace CNN Architecture: Convolution layers are independent of each other**

## 4 Evaluation

In this section, we describe the metrics within which we evaluate the effectiveness of DeepRace. Then, we outline the parameters used to train DeepRace and present the results. Moreover, we evaluate and discuss the capability of DeepRace by testing it against 60 programs which are not included in training or validation sets, 30 out of the 60 programs have data race. Furthermore, we test DeepRace using microbenchmarks available in DataraceBench [17].

### 4.1 Evaluation Metrics

To evaluate the effectiveness of DeepRace we define the following metrics:

*Precision:* The ratio of the number of data race methods classified correctly to the total number of data race predicted (either true or false).

$$Precision = \frac{Number\ of\ true\ data\ race\ predicted\ (TP)}{Number\ of\ total\ data\ race\ predicted\ (TP+FP)} \quad (1)$$

*Recall*: Measures the ratio of a number of truly predicted data race to the total number of all data races.

$$Recall = \frac{Number\ of\ true\ data\ race\ predicted\ (TP)}{Number\ of\ total\ data\ races} \quad (2)$$

*Accuracy*: Measures the ratio of correctly predicted data race and data race free files to the total number of all files.

$$Accuracy = \frac{Number\ of\ correctly\ classifed\ files(TP+TN)}{Number\ of\ total\ files(TP+TN+FP+FN)} \quad (3)$$

We also evaluate the performance of our model in detecting the lines of code that cause the data races in the method. For this purpose, we adopt the Intersect of Union method for class c and sample *t*:

$$IoU^c = Avg\ (A\ /\ ((A+B) + (C\ /\ N))) \quad (4)$$

Where *A* is the buggy code token in the source code detected by the model, *B* is the buggy code token in the source code that is not detected by the model, *C* is the not a buggy code but detected by the model as a buggy code, and *N* is all the lines of code in the method.

### 4.2 Hyper-Parameter of DeepRace CNN

Table 2 outlines the values used for different parameters of DeepRace CNN architecture. The value of these parameters is set experimentally that means by training the model several times and experimenting with different values for each parameter.

**Table 2: Hyperprameters**

| Layer | Parameters |
|---|---|
| Embedding | dimension=64 |
| Conv1 | # of filters=512, filter size=3x32 |
| Conv2 | # of filters=512, filter size=4x32 |
| Conv3 | # of filters=512, filter size=5x32 |
| MaxPool1 | shape=1x512 |
| MaxPool2 | shape=1x512 |
| MaxPool3 | shape=1x512 |
| Concatenate | shape=3, 512 |
| Flatten | shape=1x1536 |
| Dropout | rate=0.5 |
| Dense | Shape=1x2 |

### 4.3 Results

Table 3 summarizes the results according to the metrics defined in section 4.1.

In general, the best accuracies of the datasets range from 81% to 86%, which indicates DeepRace is effective in recognizing source code files containing data race from bug-free. The OpenMP dataset 2 achieved the least accuracy among others. This is due to the insufficient training samples. How well a deep model is trained and generalized is highly dependable on the size of the data set. The more training samples being available to feed to the model, the better the results could be achieved.

Figure 5 shows the accuracy rate of training and validation of DeepRace. The accuracy improves as the number of epochs increases. Setting the number epochs beyond 40 could result in slight improvement for accuracy but will also increase the training time significantly.

Additionally, to evaluate and compare our approach with other data race detectors, we examined each trained data race detector against 60 source code files which we call test set. This test set was neither included in training nor in the validation set and 30 out of the 60 files in this test set include data races. The existence of data race in OpenMP files was confirmed by *Archer* [35], a state-of-the-art data race detector for OpenMP programs and for the POSIX files, all files were analyzed by the popular tool *ThreadSanitizer* [18].

**Table 3: Results of datasets based on two architectures**

| Dataset | Precision | Recall | Accuracy |
|---|---|---|---|
| OpenMP #1 | %85 | %86 | %86 |
| OpenMP #2 | %79 | %82 | %81 |
| POSIX dataset | %81 | %83 | %83 |

Table 4 to 6 show results of DeepRace in terms of True Positive (predicting the existence of data race correctly), False Negative (incorrectly predicting a racy file as clean, so-called missing a data race), True Negative (correctly predicting a file without data race) and False Positive (predicting a clean file as having data race incorrectly). Overall, it can be observed that our approach is effective in identifying buggy and bug-free files correctly and yields a low number of a false positive and false negative.

For example, in Table 4, DeepRace only misclassified one buggy file out of 30 buggy files (1 false negative) and only predicted 2 bug-free files incorrectly as a buggy file (false positive). The results achieved by DeepRace for critical directive races in Table 5, is a bit worse than DeepRace for the private clause. As mentioned the number of source code files in the training set was not sufficient and this could be the main reason why the detector performed worse. Finally, DeepRace POSIX for lock primitives in Table 6, produced 6 false negatives and 10 false positives.

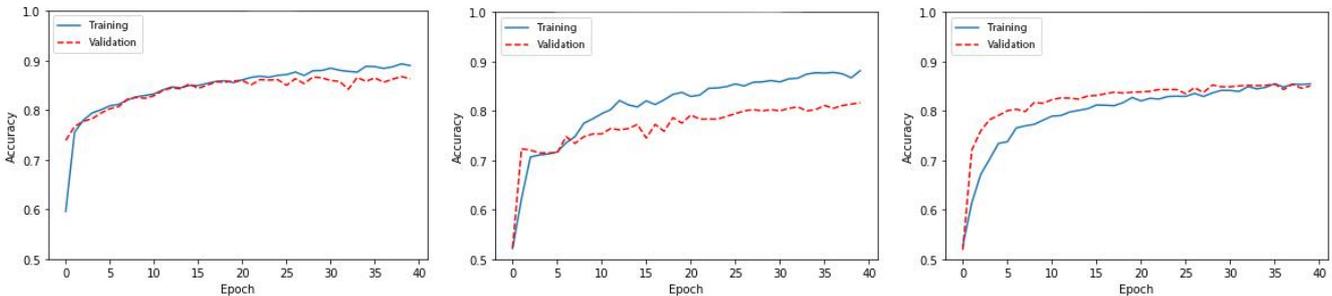

**Figure 5: Training and validation accuracy of DeepRace for three datasets based on number of epochs (OMP #1, OMP #2, POSIX)**

**Table 4: Results of DeepRace on OpenMP data races (private clause)**

|  |  | Ground Truth | |
|---|---|---|---|
|  |  | buggy | Not-buggy |
| **Predicted** | buggy | 29 | 1 |
|  | Not-buggy | 2 | 28 |

**Table 5: Results of DeepRace on OpenMP data races (critical directive)**

|  |  | Ground Truth | |
|---|---|---|---|
|  |  | buggy | Not-buggy |
| **Predicted** | buggy | 26 | 4 |
|  | Not-buggy | 1 | 29 |

**Table 6: Results of DeepRace on POSIX data races (lock primitives)**

|  |  | Ground Truth | |
|---|---|---|---|
|  |  | buggy | Not-buggy |
| **Predicted** | buggy | 24 | 6 |
|  | Not-buggy | 10 | 20 |

We also calculate the effectiveness of the DeepRace model ability to detect the lines of code that causing the bugs using the metric mentioned in section 3. We found that our model is detecting all the lines of code that causing the bug or missing only one line per method for some samples, as shown in Figure 6. However, in very few scenarios, we have noticed that the model was missing the lines that cause the problem. This show that the classifier is considering other features for classifying the file as buggy. The advantage of our model is that it shows the cases that the classifier could be misled by other features. Furthermore, we have also notice that our model is efficiently able to detect a method with several buggy lines of code at different locations in the method. This gives our model the effectiveness of handling long methods with a hundred lines of codes, as shown in Figure 7. Using the modified IoU for the buggy files, our model achieved 0.66.

We also tested DeepRace against microbenchmarks available in *DataraceBench*. *DataraceBench* is a set of OpenMP microbenchmarks with and without data race. There are a variety of data race bug patterns in these microbenchmarks. At this stage, DeepRace is capable of detecting data races based on the 3 aforementioned data race bug patterns. So, we tested DeepRace against microbenchmarks with data races similar to those 3 data race bug patterns and left detecting other data races for future work. Among microbenchmarks, only 3 of them have data race due to missing private clause. Unfortunately, no microbenchmark with data race originated from missing critical directive was available in this benchmark suite. All 3 microbenchmarks with data race have been detected as buggy by DeepRace. We also tested 17 files without data race with DeepRace and they were correctly classified as not buggy. Table 7 shows a list of files in *DataraceBench* which were analyzed by DeepRace.

Figure 8 shows file *DRB073-doall2-orig-yes.c*, which is a microbenchmark with data race in *DataraceBench*. The variable *j* in the second loop is not private and is shared among threads. This variable needs to be declared private. DeepRace shows the variable involved in the data race by highlighting the line where that variable is defined. Therefore, variable *j* is highlighted on line *4* where this variable is defined. However, since *j* and *i* variables both are defined in the same line, we can see that variable *i* is also highlighted. This means that although highlighting buggy lines is helpful in detecting bugs, reducing the granularity to words in the lines will help to gain more accuracy. The beginning of the parallel region is also highlighted by DeepRace, this shows where the

```
1   int main()
2   {
3     int i = 0;
4     omp_set_num_threads(4);
5     int *values = (int *)
6   malloc((sizeof(int)) *
7   omp_get_num_threads());
8     int threadNum = 0;
9     #pragma omp parallel shared(values)
10    {
11      threadNum = omp_get_thread_num();
12      printf("Thread %d\n", threadNum);
13      values[threadNum] =
14   doComputation((1 + threadNum) *
15   100000000);
16    }
17    for (i = 0; i < 4; i++)
18    {
19      printf("Thread %d calculated %d\n",
20   i, values[i]);
21    }
22  }
```

**Figure 6: Highlighted source code**

```
1   int main()
2   {
3     int thread_num;
4     #pragma omp parallel
5     {
6       thread_num = omp_get_thread_num();
7       …
8       #pragma omp master
9       {
10        …
11      }
12      ;
13    }
14    …
15    int a;
16    int rang;
17    #pragma omp parallel
18    {
19      rang = omp_get_thread_num();
20      a = 1;
21      #pragma omp single
22      {
23        a = 2;
24        …
25      }
26      ;
27      …
28    }
29    …
30    #pragma omp parallel
31    {
32      rang = omp_get_thread_num();
33      a = 1;
34      #pragma single copyprivate()
35      {
36        a = 2;
37      }
38      …
39    }
40  }
```

**Figure 7: Highlighted source code with several buggy lines of code**

```
1   int a[100][100];
2   int main()
3   {
4     int i,j;
5   #pragma omp parallel for
6     for (i=0;i<100;i++)
7       for (j=0;j<100;j++)
8         a[i][j]=a[i][j]+1;
9     return 0;
10  }
```

**Figure 8: Microbenchmark with data race**

parallel region begins. DeepRace was also able to detect data races in files which initially were thought to be free of these types of data race bug patterns. These files were later fixed in later versions[2].

OpenMP is mostly used by developers to parallelize *for* loops and these loops are inside the body of a method. Whereas in POSIX programs there is a main method which is run by the master thread. This master thread creates worker threads and for each of them assigns a method to run. Based on these explanations, we decided to generate ASTs for OpenMP programs on method-level, that means for each method in OpenMP program an AST is created, while for POSIX programs these ASTs are generated at file-level. This indicates that generally token vectors for POSIX programs are longer which could affect the result of training, predicting, and the time required for training. Another point needs to be considered is that the prediction is probabilistic. Consequently, there is a need for a threshold set. The higher threshold will result in less data race report which means some buggy files may be missed, whereas the lower threshold will lead to higher data race report which may increase false alarms. Here the threshold is set to 0.5 which is a common threshold for classification tasks.

It is worth mentioning that DeepRace automatically learns to distinguish between files with/without data races. This means that unlike current debugging tools and race detectors which often involve sophisticated static and dynamic analyses and algorithms, DeepRace can detect other data race patterns in source codes if appropriate training dataset for those patterns is provided.

**Efficiency**: Table 8 shows how long it takes to train DeepRace as well as how long it takes to perform validation meaning to expose DeepRace to previously unseen codes. Running training and validation on the number of files which are described in Table 1 takes less than an hour per each dataset except for POSIX dataset because of the availability of a higher number of training files and the longer length of token vectors. This table indicates that most of the time will be consumed for training DeepRace, and once DeepRace is trained, deploying it for new source code files for debugging purposes will not take much time as one may require with dynamic race detectors to execute the programs.

**Discussion**: Our experiments with DeepRace were conducted using a corpus of C files collected from GitHub. For each C file, an AST was generated and then nodes of that AST were extracted to create a token vector. To keep the ASTs small, we downloaded small size files from GitHub. This might cause some biases in the dataset. In the future, we plan to collect more files especially larger. Moreover, the information that is extracted from ASTs can be improved. For instance, we can also include the relations between the nodes. We believe adding these steps will further improve the results achieved by DeepRace. From another aspect, in this research, 3 data race bug patterns were targeted. These data race bug patterns cannot and will not fully represent all data race types in multi-threaded C programs. Analyzing more sophisticated patterns and creating datasets accordingly is planned to conduct in the future. Finally, not all C parsers produce the same ASTs, experimenting with disparate parsers might lead to higher or lower accuracies.

## 5 Conclusion

In this paper, we propose an approach to predict data races in source codes via deep learning. We leverage the power of the convolutional neural network to train DeepRace, a data race detector which can predict whether a source code file contains a data race or not, we expand the approach further that it is able to highlight the lines of codes which are involved in the data race. As the experimental results confirm, the trained DeepRace is efficient in classifying buggy or clean source code correctly comparable to the state-of-the-art tools, achieving accuracies between 81% and 86%. DeepRace automatically learns to discriminate between buggy and bug-free source code. Considering the effort and cost of developing conventional data race detectors with sophisticated algorithms, building and training DeepRace for identifying other patterns of data races is more convenient and feasible and does not require designing complex algorithms or code analysis. Furthermore, the DeepRace model was efficiently able to detect all the lines of code that causing the bugs or missing only one line for some samples. However, in very few scenarios, we have noticed that the model was missing the lines that causing the problem, we leave these cases as future work.

**Table 7: List of microbenchmarks analyzed by DeepRace**

| # | File name | Groundtruth | DeepRace Result |
|---|---|---|---|
| 1 | DRB020-privatemissing-var-yes | Data race | Data race |
| 2 | DRB028-privatemissing-orig-yes | Data race | Data race |
| 3 | DRB073-doall2-orig-yes | Data race | Data race |
| 4 | DRB041-3mm-parallel-no | No data race | No data race |
| 5 | DRB042-3mm-tile-no | No data race | No data race |
| 6 | DRB043-adi-parallel-no | No data race | No data race |
| 7 | DRB044-adi-tile-no | No data race | No data race |
| 8 | DRB046-doall2-orig-no | No data race | No data race |
| 9 | DRB055-jacobi2d-parallel-no | No data race | No data race |
| 10 | DRB056-jacobi2d-tile-no | No data race | No data race |
| 11 | DRB057-jacobiinitialize-orig-no | No data race | No data race |
| 12 | DRB058-jacobikernel-orig-no | No data race | No data race |
| 13 | DRB059-lastprivate-orig-no | No data race | No data race |
| 14 | DRB060-matrixmultiply-orig-no | No data race | No data race |
| 15 | DRB061-matrixvector1-orig-no | No data race | No data race |
| 16 | DRB063-outeronly1-orig-no | No data race | No data race |
| 17 | DRB064-outeronly2-orig-no | No data race | No data race |
| 18 | DRB065-pireduction-orig-no | No data race | No data race |
| 19 | DRB067-restrictpointer1-orig-no | No data race | No data race |
| 20 | DRB076-flush-orig-no | No data race | No data race |

**Table 8: Time required for training and validation of a data race detector (min:sec:ms)**

| Dataset | Training | Validation |
|---|---|---|
| OpenMP Dataset 1 | 06:35:78 | 00:00:35 |
| OpenMP Dataset 2 | 01:50:59 | 00:00:13 |
| POSIX Dataset | 23:52:06 | 00:00:86 |

---

[2] https://github.com/LLNL/dataracebench/issues/1